\documentclass{jfm}
\usepackage{natbib}
\usepackage{graphicx}
\usepackage{epstopdf, epsfig}
\usepackage{caption}
\usepackage{subcaption}
\usepackage{gensymb}
\usepackage{color}
\usepackage{amsmath,mathtools}
\usepackage{amsfonts}
\usepackage{amssymb}
\usepackage{amsmath} 
\usepackage{upgreek}
\usepackage{mathrsfs}
\usepackage{float}
\usepackage{hyperref}
\usepackage[normalem]{ulem}

\newcommand{\be}{\begin{equation}}
\newcommand{\ee}{\end{equation}}

\newcommand{\We}{W\kern-0.15em e\,}
\newcommand{\Bo}{B\kern-0.05em o\,}
\newcommand{\Oh}{O\kern-0.05em h\,}
\newcommand{\R}{R\kern-0.15em e\,}

\newdimen\figrasterwd
\figrasterwd\textwidth

\shorttitle{Water entry of small hydrophobic spheres on an axisymmetric wavefield}
\shortauthor{C.T. Gabbard and others}

\title{Water entry of small hydrophobic spheres on an axisymmetric wavefield}

\author{Chase T. Gabbard$^{1,2, \corresp{\email{chase.gabbard@me.gatech.edu}}}$, Mario Ibrahim$^{1,3}$, Joseph Quinton$^{1}$, Eli Silver$^{1}$, Jesse Belden$^{4}$, \and Daniel M. Harris$^{1, \corresp{\email{daniel\_harris3@brown.edu}}}$}

\affiliation{$^{1}$School of Engineering, Brown University, Providence, RI 02912, USA  \\ $^{2}$George W. Woodruff School of Mechanical Engineering, Georgia Institute of Technology, Atlanta, GA 30332, USA \\ $^{3}$Institute of Mechanical Engineering, École Polytechnique Fédérale de Lausanne (EPFL), 1015 Lausanne, Switzerland \\ $^{4}$Naval Undersea Warfare Center, 1176 Howell Street, Newport, RI 02841, USA}

\begin{document}

\maketitle

\begin{abstract}
When a sphere crosses an air-water interface it can entrain a significant volume of air, a process relevant to numerous naval, industrial, and environmental settings. While air entrainment through sphere impact onto quiescent interfaces has been extensively studied, real-world interfaces are inherently unsteady, and the influence of surface waves is less understood. In this work, we systematically investigate the effect of waves on air entrainment in a highly controlled setting: small hydrophobic spheres impacting an axisymmetric standing wavefield. By analyzing the resulting cavity across a wide parameter space, including wave phase, driving amplitude, and frequency, we reveal that local interface deformation dramatically alters the maximum cavity volume. This effect is driven by a geometric modulation of the splash curtain, which shifts the transition between cavity closure modes. We demonstrate that the influence of waves in our setting is well parameterized by the instantaneous interface slope evaluated at the radius of the sphere, successfully collapsing our experimental data alongside the traditional Weber and Bond numbers. 
\end{abstract}

\begin{keywords}
multiphase flow, wave-structure interactions
\end{keywords}

\section{Introduction}\label{sec:introduction}

When a solid object impacts an air–water interface, an air cavity can form behind the object above a critical impact velocity that depends on the object's wettability \citep{duez2007making}, roughness \citep{worthington1900iv,zhao2014wetting}, temperature \citep{marston2012cavity}, surface contamination \citep{worthington1900iv}, as well as interfacial features such as bubbles, oil layers, or floating material \citep{speirs2018entry,watson2018jet,smolka2019sphere,watson2024compound}. Understanding the physics governing transient air cavities has driven over a century of research \citep{truscott2014water}, dating back to the seminal observations of \citet{worthington1897v,worthington1900iv}. This interest is sustained by the ubiquity of impact-driven air cavities found in anthropogenic applications, such as naval structure design \citep{von1929impact_modified,gilbarg1948influence,may1975water} and drag-reduction techniques \citep{vakarelski2011drag}; and nature, like diving birds and beetles \citep{ropert2004between,miller2016diving}, and biolocomotion at natural air--water interfaces \citep{glasheen1996hydrodynamic,bush2006walking}. In such settings, the interface is inherently wavy, yet how perturbed interfaces influence air entrainment remains largely unexplored.

The impact of a hydrophobic sphere with radius $R$ and velocity $U$ onto a liquid bath (density $\rho$; surface tension $\sigma$) entrains a maximum volume of air $V_\textrm{m}$ that depends on the coupled evolution of the cavity and splash curtain. Cavity type is categorized based on the mechanism driving its closure. As outlined by \citet{aristoff2009water}, the canonical cavity types are quasi-static seal, shallow seal, deep seal, and surface seal. The dimensionless parameters governing the transition between cavity types for a still bath are the Weber number $\We=\rho U^{2} R / \sigma$ and Bond number $\Bo=\rho g R^{2} /\sigma$, which compare the influence of gravity and inertia to capillarity. At high $\We$, inertia drives rapid cavity growth and the closure type transitions from deep seal to surface seal. While deep seals result in a primary submarine pinchoff driven mainly by hydrostatic pressure, surface seals result from surface tension and aerodynamic pressure difference acting together to recombine the splash crown above the impact location prior to submarine pinchoff (cf. Figure~\ref{Fig1}($d,e$)) \citep{gilbarg1948influence,duclaux2007dynamics,aristoff2009water,marston2016crown,eshraghi2019seal}. This transition limits the maximum volume of air in the cavity, and the positive relationship between $V_\textrm{m}$ and $\We$ for deep seals reverses for surface seals \citep{mansoor2014water}. Consequently, cavity volume is a non-monotonic function of $\We$ that peaks at the $\We$ separating deep and surface seals.

Increasing the cavity volume beyond this maximum requires extending the $\We$ range of deep seal cavities. At large $\Bo$, where surface sealing is driven primarily by aerodynamic pressure, reducing the ambient gas pressure extends the deep seal regime to higher $\We$ \citep{gilbarg1948influence,may1968cavity,marston2016crown}. Alternatively, surface seals can be suppressed by positioning a splash guard above the impact site to prevent the splash curtain from recombining \citep{mansoor2014water}. However these interventions are limited to laboratory settings. Under typical real-world conditions, surface seal is more likely suppressed due to imperfections in the surface of the sphere \citep{gilbarg1948influence}, a liquid pre-wetting the sphere \citep{may1968cavity}, or material floating atop the bath \citep{watson2018jet,watson2024compound}. The influence of waves on closure type and cavity evolution remains less explored, despite their known effects on rigid-body slamming forces \citep{moalemi2023cylinder,zhao2023vertical} and drop-impact dynamics \citep{siscoe1971water}. \citet{zhao2023vertical} took a first step toward addressing this gap by examining the cavities of centimeter-scale spheres impacting a traveling plane wave. Their non-axisymmetric scenario naturally induces non-axisymmetric fluid dynamics, evident in the splash crown and air cavity evolution. For impacts at a planar wave crest, they observed an increased crown diameter and delayed pinch-off time relative to the still-water case, however their experiments were limited to a single sphere size, impact velocity, and wave frequency, making it difficult to draw broader conclusions.  In order to build a firm fundamental foundation, we focus herein on a fully axisymmetric setting and explore a much broader range of wave and impact parameters.  Our systematic approach enables a mechanistic interpretation of our key findings, which informs a single geometric wave parameter that collapses our air cavity measurements at each $\We$ and $\Bo$.

Here, we show that surface waves can dramatically alter the volume of the air cavity formed during water entry. This effect is driven by the instantaneous interface geometry at the location of impact, which modulates the transition between the canonical deep seal and surface seal regimes for a given $\We$ and $\Bo$. As a consequence, our system isolates the role of surface geometry on air entrainment from other bulk-flow effects such as non-axisymmetric flows \citep{zhao2023vertical} or rotational effects \citep{yi2021water}.

\section{Experiment}\label{sec:experiment}

\begin{figure}
\centering
\includegraphics[width=\textwidth]{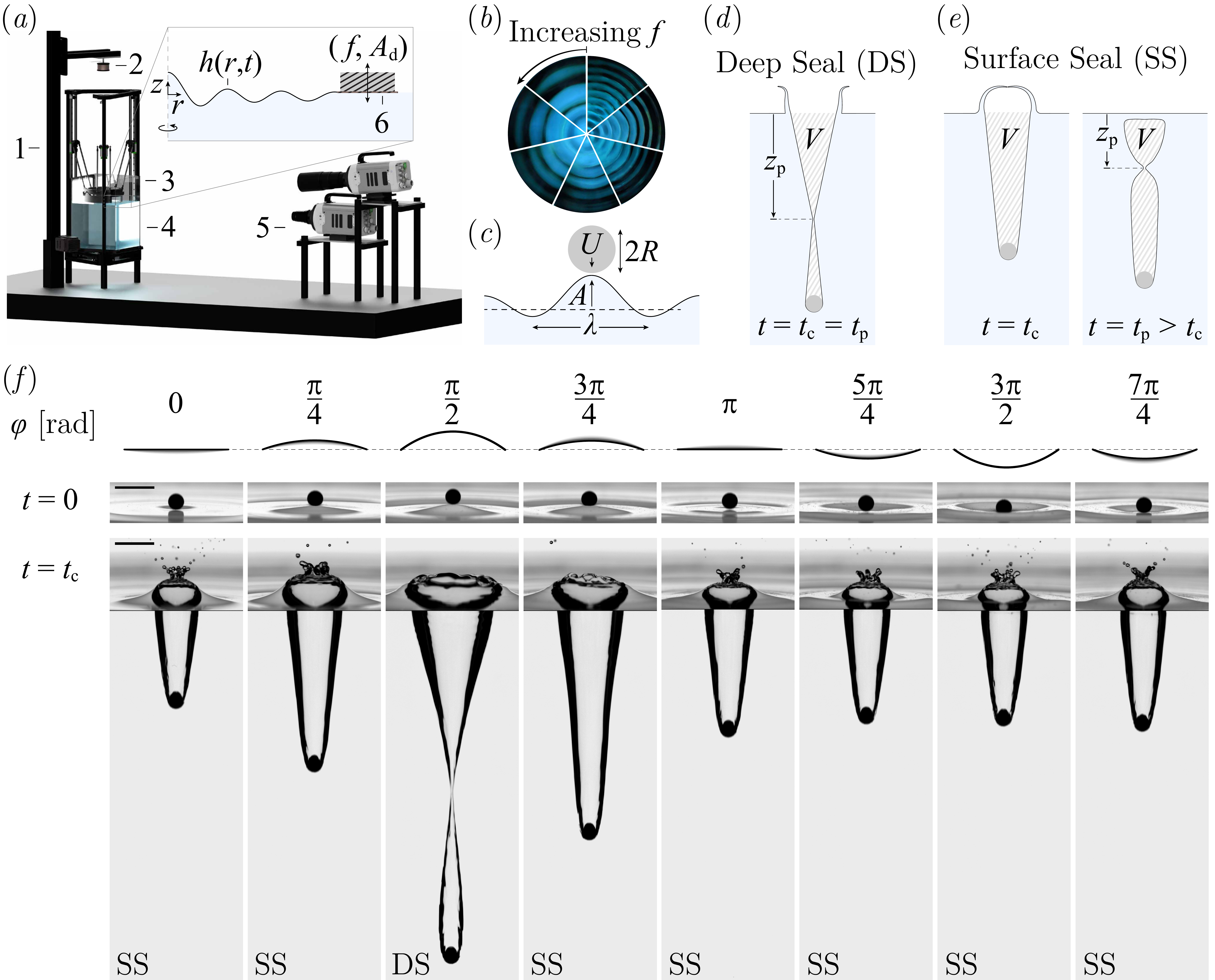} \caption{($a$) Schematic of the experimental setup, including the (1) drop tower, (2) electromagnet, (3) heaving cylindrical wavemaker (Appendix~\ref{appA}), (4) water tank, and (5) high-speed cameras. The inset schematic depicts the axisymmetric wavefield generated by the heaving motion of (6) a cylindrical ring at the air-water interface. ($b$) Top-view of the wavefields produced within the 120 mm ring for seven frequencies ranging from 6.79 to 16.54 Hz, where reflected light reveals the nodes and antinodes as dark and bright regions, respectively \citep{shao2021role}. ($c$) Schematic of a sphere at the moment of impact on a wave crest, with the governing parameters labeled. Following impact, the air cavity grows until it closes at $t=t_\textrm{c}$ either below the initial water level for a ($d$) deep seal (DS), or above it for a ($e$) surface seal (SS). The submarine pinchoff depth $z_\textrm{p}$ occurs at time $t_\textrm{p}$, and the hatched regions indicate the cavity volume $V$. ($f$) Columns showing the wave phase $\varphi$ definition (top), moment of impact (middle), and the resulting air cavity at $t=t_\textrm{c}$ for $\We=560$, $\Bo=0.54$, $f=9.64$ Hz, and $A_\textrm{d}=0.1$ mm (Movie 1). Scale bars are 10 mm.} \label{Fig1}
\end{figure}

Our experimental setup (Figure~\ref{Fig1}($a$)) features a motorized drop tower (1) supporting an electromagnet (2), and a custom heaving wavemaker (3) that generates an axisymmetric wavefield on a deep water bath (4) with density $\rho=997$ kg/m$^3$ and surface tension $\sigma=71.3$ mN/m, and dimensions 200 mm $\times$ 200 mm $\times$ 195 mm (L $\times$ W $\times$ H). Two synchronized Photron Nova R5 high-speed cameras (5) capture the splash dynamics and submarine cavity evolution at 7,500 fps using a Nikon 200 mm Nikkor and Venus Optics Laowa 100 mm Ultra Macro lens, respectively. Axisymmetric standing waves are formed by vertically driving a cylindrical ring—labeled (6) in the Figure~$\ref{Fig1}$($a$) inset—at a prescribed frequency $f$ and amplitude $A_\textrm{d}$ using three motorized axes. The ring has an inner radius of 65 mm and an outer radius of 75 mm. The full span of operating conditions is 6.79 Hz $\le f \le$ 16.54 Hz and 0.04 mm $\le A_\textrm{d} \le$ 0.16 mm which produce axisymmetric wave patterns like those shown in Figure~\ref{Fig1}($b$). Additional details about the wavemaker can be found in Appendix~\ref{appA}, and motivation for the particular frequencies explored is provided in Sec.~I of the Supplemental Information.  

In each experiment, a hydrophobic sphere of radius $R=1.98, 2.78$ mm ($\Bo=0.54,1.06$) is released from an electromagnet at a height $H_0$ above the bath. The sphere impacts the wavefield at the axis of symmetry $r=0$ with velocity $U \approx \sqrt{2gH_0}$, such that $54 \leq We \le 871$. The spheres are 440C stainless steel ($\rho_s$ = 7665 kg/m$^3$) and rendered hydrophobic using a superhydrophobic spray coating (Rust-Oleum NeverWet) following the protocol detailed in \citet{galeano2021capillary}. Once coated, spheres always formed an air cavity. To minimize the influence of possible degradation of the coating, each sphere was used no more than three times. The moment of impact is schematically shown in Figure~\ref{Fig1}($c$). The forcing frequency $f$ defines the wavelength $\lambda$ via the deep-water gravity-capillary wave dispersion relation, giving a range $1.4 \ \mathrm{cm} \le \lambda \le 4.0 \ \mathrm{cm}$. The central amplitude $A$ at the moment of impact is dictated by both the drive amplitude $A_\textrm{d}$ and the impact phase $\varphi$, the latter of which is controlled by adjusting the relative timing between the wavemaker motion and sphere release.

The high-speed recordings of the splash and cavity evolution are analyzed using a custom image processing script in MATLAB (see Sec.~II in the Supplemental Information). Cavity volume $V(t)$ is computed frame-by-frame as a volume of revolution between the undisturbed surface and the bottom of the sphere. For both deep seal (DS) and surface seal (SS), $V(t)$ is tracked up to the first submarine pinchoff time $t_\textrm{p}$, which occurs after the cavity closure time $t_\textrm{c}$ for SS, as shown in Figure~\ref{Fig1}($d,e$). Then, the maximum volume is $V_\textrm{m}=\textrm{max}(V(t))$. We also measure the pinchoff depth $z_\textrm{p}$ relative to the undisturbed free surface. We performed each experiment three times, and the plot markers and uncertainty bars used in this study correspond to the average and standard deviation across those trials.

The first row in Figure~\ref{Fig1}($f$) schematically defines the wave profile for each phase $\varphi$. For a given $A_\textrm{d}$, the maximum and minimum central wave height $A$ occur at $\varphi=\pi/2$ and $3\pi/2$, respectively. The panels in the rows below show the initial sphere impact ($t=0$, middle row) and the final cavity and splash shape at closure ($t=t_\textrm{c}$, bottom row) for fixed impact conditions ($\We= 560$, $\Bo=0.54$, $f=9.64\text{ Hz}$, $A_\textrm{d}=0.1\text{ mm}$). For wave phases with convex-upward curvature ($\pi/4 - 3\pi/4$), the cavity size and the sphere depth at closure increase. Most notably, at $\varphi=\pi/2$ (corresponding to maximum $A$) the cavity length has more than doubled relative to the same impact on a quiescent bath, and the closure type has changed to deep seal. In contrast, for waves with concave-upward curvature ($5\pi/4 - 7\pi/4$), the cavity size is less sensitive to the interface geometry at this $\We$ and $\Bo$, and the cavities close through a similar surface seal in each case for these parameters. Visualizations similar to Figure~\ref{Fig1}($f$) for increasing frequency $f$ and driving amplitude $A_\textrm{d}$ are included in Sec.~III of the Supplemental Information.

\section{Results} \label{sec:results}

\begin{figure}
\centering
\includegraphics[width=0.5\textwidth]{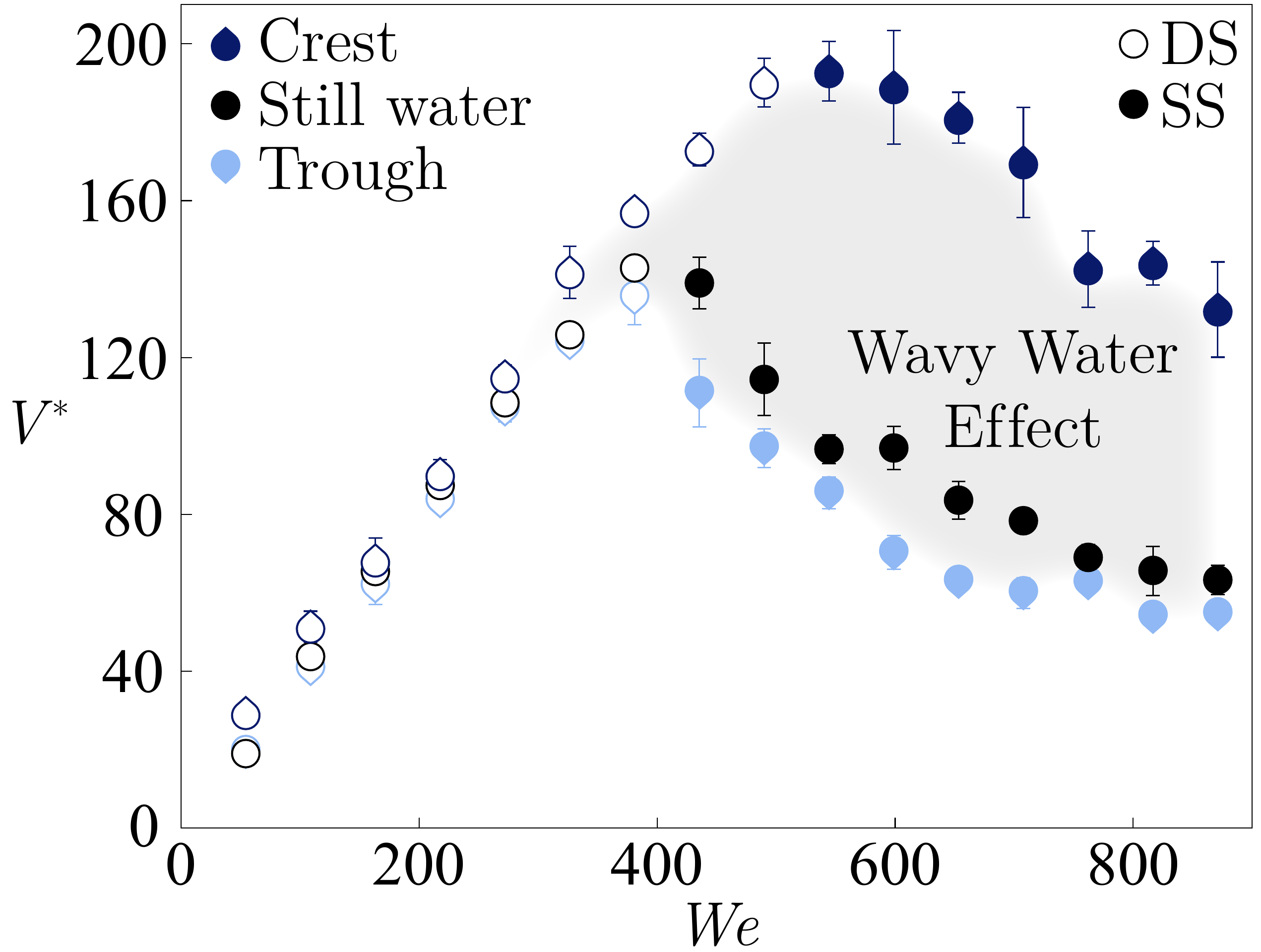} \caption{Dimensionless maximum cavity volume $V^*=V_\textrm{m}/V_\textrm{s}$ against Weber number $\We$ for impacts on still water (black circles; Movie 2), a wave crest (dark blue upward cusped circles; Movie 3), and a trough (light blue downward cusped circles; Movie 4) for $A_{\textrm{d}}=0.1$ mm, $f=9.64$ Hz, and $\Bo=0.54$. Open and filled markers indicate deep seal (DS) and surface seal (SS), respectively. The shaded region highlights the Wavy Water Effect (WWE) for this set of parameters.} \label{Fig2}
\end{figure}

Since surface waves can modify the cavity closure type (Figure~\ref{Fig1}($f$)), we expect wave properties to influence cavity volume most sensitively near the transition between deep seal and surface seal. Figure~\ref{Fig2} plots the dimensionless maximum cavity volume $V^*=V_\textrm{m}/V_\textrm{s}$ (where $V_\textrm{s}=4/3 \, \pi R^3$) against $\We$ for impacts on still water (Movie 2), wave crests $\varphi=\pi/2$ (Movie 3) and wave troughs $\varphi=3\pi/2$ (Movie 4) for $A_\textrm{d}=0.1$ mm, $f=9.64$ Hz, and $\Bo=0.54$. The open and filled markers show whether the cavity closed via a deep seal or surface seal, respectively. At low $\We$, only deep seals occur, with crest impacts slightly increasing the cavity volume, and trough impacts decreasing the volume. Other than this offset, cavity volume increases at a similar rate with $\We$ across all deep-seal conditions. However, above the critical threshold for still-water surface sealing ($\We\approx400$), the cavity volume trends diverge and depend much more sensitively on the interface shape. For crest impacts, the transition to surface seal is notably delayed, allowing $V^*$ to increase up to $\We\approx550$. In contrast, the still-water and trough impacts transition to surface seal at $\We\approx400$, above which increasing $\We$ further decreases $V^*$. Notably, crest impacts induce a larger deviation from the still-water reference than trough impacts for this set of wave parameters, mirroring the qualitative asymmetries in the influence of $\varphi$ shown in Figure~\ref{Fig1}($f$). The cumulative span in $V^*$ at a fixed Weber number is notable, more than doubling at $\We\approx550$ as further illustrated in Sec.~IV of the Supplemental Information. This sensitivity to waves demonstrates what we define here as the Wavy Water Effect (WWE). The submarine pinchoff depth $z_\textrm{p}$ and time $t_\textrm{p}$ are also influenced by the WWE, as shown in Appendix~\ref{appB} (Figure~\ref{Fig7}); a parallel analysis of all pinchoff metrics is provided in Sec.~V of the Supplemental Information.

\begin{figure}
\centering
\includegraphics[width=\textwidth]{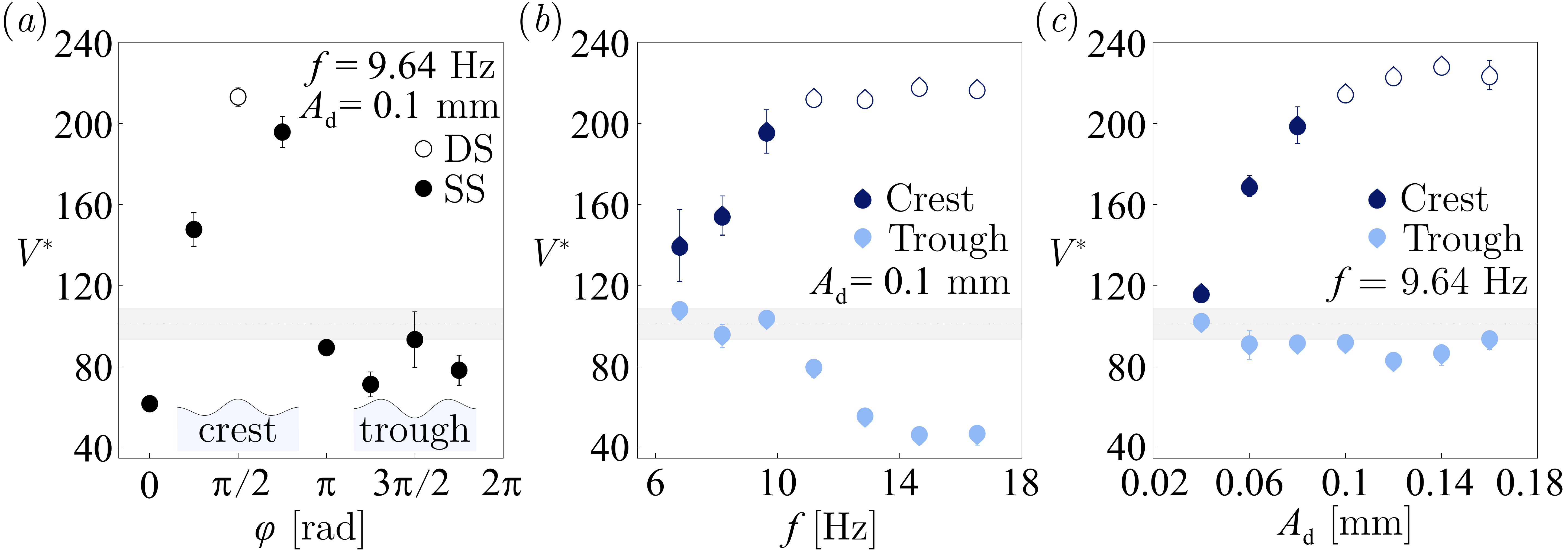} \caption{Dimensionless maximum cavity volume $V^*=V_\textrm{m}/V_\textrm{s}$ plotted against ($a$) wave phase $\varphi$, ($b$) driving frequency $f$, and ($c$) driving amplitude $A_{\textrm{d}}$ for $\We=560$ and $\Bo=0.54$ (Movies 1, 5-8). Open and filled symbols indicate a deep seal (DS) and surface seal (SS), respectively. Baseline $V^*$ for impact on still water is shown by the dashed line and shaded uncertainty band.} \label{Fig3}
\end{figure}

For a fixed $\We$ and $\Bo$, the cavity volume depends on our three wave control parameters: phase $\varphi$, frequency $f$, and driving amplitude $A_\textrm{d}$. By independently varying each parameter, we show how they influence $V^*$ in Figure~\ref{Fig3}, while also making comparison to the still-water control (dashed black line). Additionally, we include plots of the measured non-dimensional wave amplitude $A^*=A/R$ versus each parameter in Sec.~VI of the Supplemental Information. Figure~\ref{Fig3}($a$) shows that while $V^*$ rises and falls in phase with $\varphi$, its influence is strongest around $\pi/2$ (crest) with a lesser effect near $3 \pi/2$ (trough). Figure~\ref{Fig3}($b$) shows that $f$ has an opposite effect on crest (dark blue) and trough (light blue) impacts. As $f$ increases, the measured wave amplitude $A$ remains nearly constant (Figure~S7) but the wavelength $\lambda$ shortens, and progressively amplifies air entrainment for crest impacts while further suppressing it for troughs. For crest impacts, this frequency-driven enhancement triggers a transition to the deep-seal regime, beyond which $V^*$ plateaus. Figure~\ref{Fig3}($c$) shows a similar trend as driving amplitude $A_\textrm{d}$ (and thereby $A$) is increased. In particular, we see a similar increase in $V^*$ and transition to deep seal as $A_\textrm{d}$ increases for crest impacts. Conversely, cavity size is less sensitive to $A_\textrm{d}$ for trough impacts, only slightly decreasing relative to the still-water baseline. These isolated trends highlight the complementary roles of phase, frequency, and driving amplitude, and motivate the need for a comprehensive framework capable of reconciling the trends across the entire multi-dimensional parameter space.

In an attempt to rationalize the various parametric dependencies revealed in the last section, as well as isolate the mechanism by which surface waves modulate maximum cavity volume, we first ask whether surface waves serve to primarily modify the submarine cavity shape or the trajectory of the splash curtain. In our experiments, the splash curtain emerges on the timescale of the sphere passing through the interface ($R/U$), which is two orders of magnitude faster than the wave oscillation period ($1/f$). As such, the interface appears frozen during the emergence of the ejecta sheet, with the instantaneous geometry of the interface therein defining the initial conditions for its ultimate dynamic evolution.

\begin{figure}
\centering
\includegraphics[width=\textwidth]{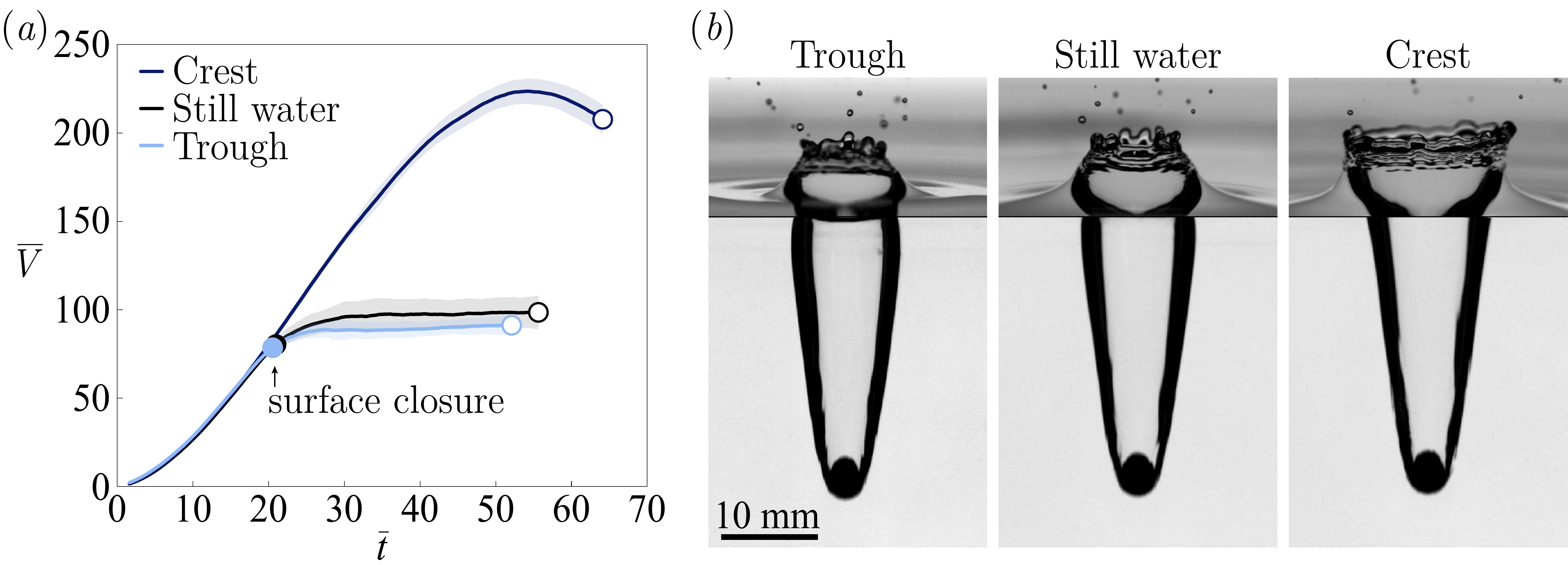} \caption{($a$) Dimensionless cavity volume $\overline{V}=V(t)/V_\textrm{s}$ against dimensionless time since impact $\overline{t}=\left(t-t_0\right)U/R$ for a hydrophobic sphere ($\Bo=0.54$, $\We=560$) impacting a wave crest (dark blue), trough (light blue), and still water (black). Wave parameters are $f=9.64$ Hz and $A_\textrm{d}=0.16$ mm. Solid lines and shaded regions represent the average and standard deviation across three repeat trials. Filled and open circles denote surface closure and submarine pinchoff, respectively. Note that surface closure does not occur for the crest case. ($b$) Splash crown and cavity profiles at $\overline{t}=15$ for the conditions in ($a$).}
\label{Fig4}
\end{figure}

For fixed impact parameters, the early stages of cavity growth are largely unaffected by this surface waviness. Figure~\ref{Fig4}($a$) plots the dimensionless cavity volume $\overline{V}=V(t)/V_\textrm{s}$ against dimensionless time since impact $\overline{t}=(t-t_0) \, U/R$ for a hydrophobic sphere ($\Bo=0.54$, $\We=560$) impacting a wave crest, trough, and still water. For $\overline{t}\lesssim20$, the cavity volume grows nearly identically across all three cases. Figure~\ref{Fig4}($b$) shows a side-by-side comparison of these impacts for $\overline{t}=15$, prior to the surface seal appreciably altering the cavity growth. The uniformity of the cavity profiles across these disparate impact conditions confirms a mostly wave-agnostic sub-surface growth phase. This decoupling may allow established theoretical models for transient cavity shapes in calm water \citep{duclaux2007dynamics, aristoff2009water} to be directly applied to wavy-water impacts for timescales prior to surface closure ($t < t_\textrm{c}$).

Minor differences in the cavity profile are observed only visible near the free surface. For a planar capillary-gravity wave, the amplitude of the wave-induced pressure fluctuation (sometimes referred to as the perturbation pressure \citep{kundu2024fluid}) is given by:  
\begin{equation}
    P_w(z) = A_0\left( \rho g + \frac{4 \pi^{2}\sigma}{ \lambda^2} \right) e^{\frac{2 \pi z}{\lambda}}.
\end{equation}
Thus, the influence of the waves on the pressure distribution in the fluid is expected to be isolated to the near-surface region due to the exponential decay with depth. Setting this pressure equal to the baseline hydrostatic pressure ($P_g(z) = \rho g z$) allows us to determine a crossover depth $z=-d_\textrm{c}$ beyond which the hydrostatic pressure dominates. Considering the full range of experimental parameters in our study, we can arrive at an upper-bound estimate of $d_\textrm{c}/{R} \lesssim 1.0$. Therefore, consistent with our observations, the wavy interface is expected to modify the cavity profile through the surrounding pressure field only over depths on the order of the sphere radius, \(z\sim R\). A similar near-surface modification of cavity shape has been reported for water entry into a rotating fluid bath \citep{yi2021water}.

In contrast to the nearly identical sub-surface cavities, the influence of surface waves on the splash curtains is immediately apparent, and the trajectories of $\overline{V}$ diverge rapidly once surface seal occurs. While the splash curtains for both the trough and no-wave impacts are already accelerating inward to seal the surface just beyond $\overline{t}=20$ (stopping airflow into the cavity and causing the volume to plateau), the splash crown in the crest case continues to expand vertically and radially outward, consistent with measurements for impacts on the crest of traveling waves \citep{zhao2023vertical}. Consequently, impact on a wave crest delays surface seal, prolonging the growth phase of the underlying air cavity such that it ultimately achieves a maximum volume more than $200\%$ larger than that of the corresponding trough and still-water impacts. Conversely, impact on a trough accelerates surface seal relative to still-water impacts, though this effect is comparatively weaker for these parameters. The splash crown is thus the component most sensitive to surface waves, and altering its evolution, rather than the cavity expansion rate, is the primary mechanism driving the changes in the maximum cavity volume reported in our study. This decoupling of the sealing timescale from the core cavity expansion mirrors the mechanism by which reduced ambient pressure alters water-entry dynamics \citep{gilbarg1948influence}.

\begin{figure}
\centering
\includegraphics[width=1\textwidth]{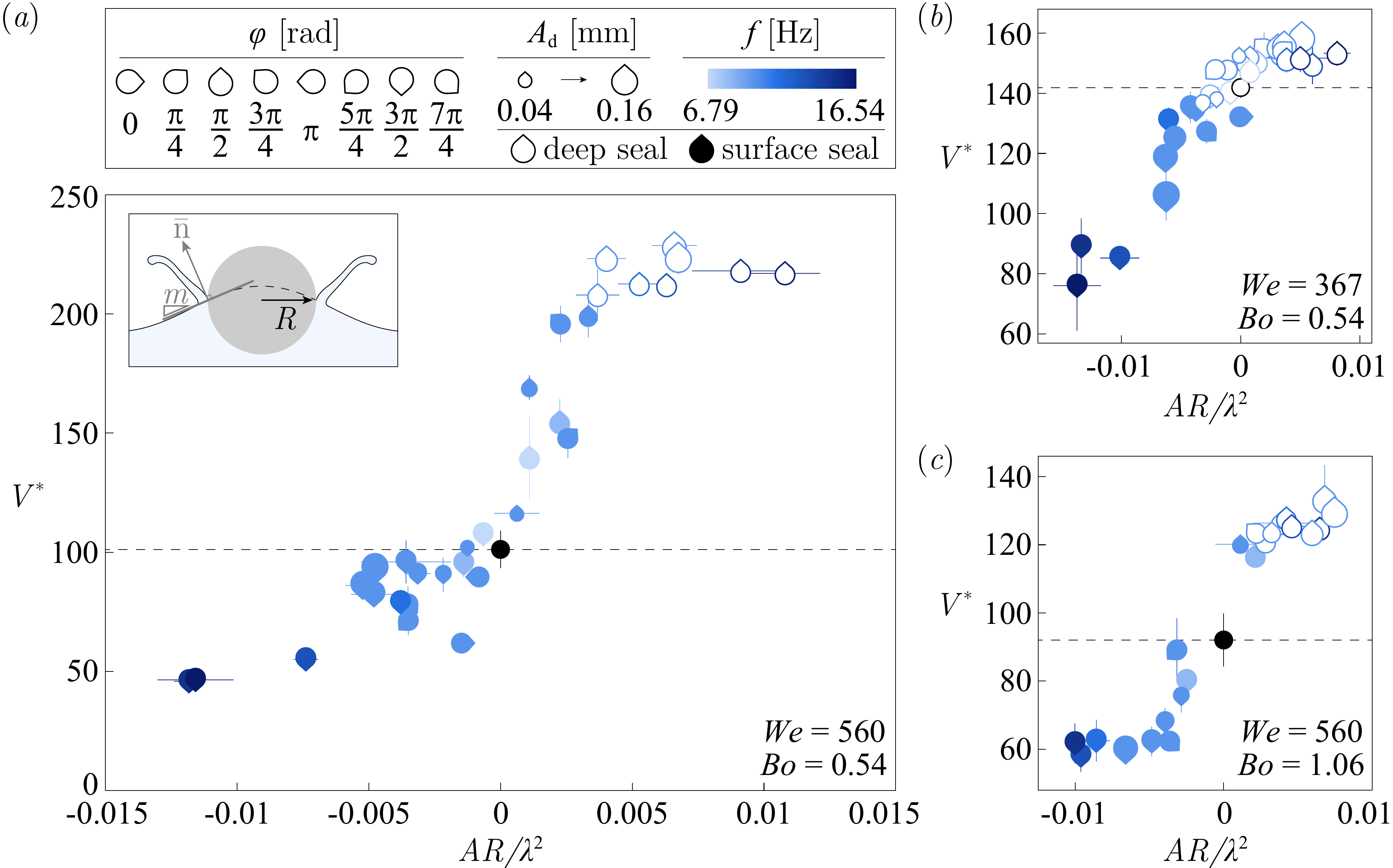} \caption{Dimensionless maximum cavity volume $V^*=V_\textrm{m}/V_\textrm{s}$ against the characteristic wave slope $m \sim AR/\lambda^2$ at $r=R$ (see inset) for ($a$) $\We=560$ and $\Bo=0.54$, ($b$) $\We=367$ and $\Bo=0.54$, and ($c$) $\We=560$ and $\Bo=1.06$. The marker shape, size, and color reflect the wave phase $\varphi$, driving amplitude $A_\textrm{d}$, and driving frequency $f$, respectively, as defined in the legend. Open and filled markers denote deep seal and surface seal, respectively. Black circles show impacts on still water for identical conditions.} \label{Fig5}
\end{figure}

To parameterize the influence of the interface geometry on the splash curtain ejection, we hypothesize that the surface slope at the scale of the impacting body plays a dominant role in defining the outcome (see Figure~\ref{Fig5}($a$) inset). Modeling the surface topology as an axisymmetric standing wavefield centered about the impact axis, $h(r,t) = A_0 J_0(kr)\sin(\omega t+\varphi)$, the surface slope is then given by
\begin{equation} \label{eq1}
\frac{\partial h}{\partial r} = -A_0k J_1(kr)\sin(\omega t+\varphi)
\end{equation}
where $k=2\pi/\lambda$ is the wavenumber and $\omega=2\pi f$ is the angular frequency. For a given impact phase $\varphi$ at $t=0$ (such that $A=A_0\sin{\varphi}$), one can then estimate the characteristic interface slope $m$ at the sphere radius $R$ (assuming $R \ll \lambda$) as
\begin{equation}
m \sim \frac{A R}{\lambda^2}.
\end{equation}
This parameter is inherently non-dimensional, and can also be interpreted as the ratio of the sphere radius ($R$) to the interface radius of curvature ($\sim 1/Ak^2$) at the impact site.  

Figure~\ref{Fig5}($a$) shows the dimensionless maximum cavity volume $V^*=V_\textrm{m}/V_\textrm{s}$ as a function of the local wave slope $m \sim A R / \lambda^2$ across all wave phases $\varphi$ (marker orientation), driving amplitudes $A_\textrm{d}$ (marker size), and frequencies $f$ (marker color) for $\We = 560$ and $\Bo = 0.54$. The data collapse onto a single curve: $V^*$ increases monotonically with $A R / \lambda^2$, passes through the still-water reference (dashed line), and ultimately plateaus at high wave slopes. This plateau coincides with a transition in the closure mechanism to a deep seal, wherein the influence of surface geometry does not further affect cavity growth. Figure~\ref{Fig5}($b$) and \ref{Fig5}($c$) present similar plots but for a different $\We$ and $\Bo$, respectively. Because $\We$ or $\Bo$ dictate the reference still-water cavity volume $V^*$ and closure type, changing these parameters alters how sensitive the final cavity volume is to the impact phase. For instance, in the still-water DS regime (Figure~\ref{Fig5}($b$)), impact on a trough can induce SS, and dramatically reduce the volume.  In each case, the data collapse onto a single trend across all control parameters. Similar collapse is observed for the submarine pinchoff depth $z_\textrm{p}$ and time $t_\textrm{p}$, as shown in Appendix~\ref{appC} (Figure~\ref{Fig8}). Consequently, we have demonstrated that the influence of axisymmetric surface waves is entirely captured by the local wave slope $m$, which alongside $\We$ and $\Bo$ fully dictates the maximum cavity volume during the water entry of hydrophobic spheres. Building off of the success of the proposed wave parameterization herein, we further hypothesize that high-speed ($fR/U \ll 1$) impacts on a 3D interface with long waves ($R/\lambda\ll 1$) could be characterized by three local geometric parameters: two principal non-dimensional interface curvatures and a mean surface slope. This hypothesis could be readily evaluated in future work.

\section{Discussion} \label{sec:discussionandconclusion}

Although established here for centered impacts on an axisymmetric wavefield, the local wave slope $m$ introduces a single unifying parameter motivated by a clear physical mechanism, and may be adaptable to more complex interfaces and other solid-liquid impact scenarios. Because air cavities form during high-speed water entry regardless of surface wettability \citep{duez2007making}, these findings are broadly applicable. Potential applications include impacts on highly curved static interfaces, such as particulate capture by raindrops \citep{speirs2023capture}, and droplet splashing on perturbed liquids \citep{siscoe1971water}. In fact, similar geometric scaling has proven successful for rationalizing the non-axisymmetric scenario of an impacting droplet splashing on a slowly moving bath \citep{sykes2025fast}, suggesting that our framework might similarly extend to drops striking wavy surfaces. While impactor motion on timescales comparable to cavity formation can dramatically alter cavity shape \citep{aristoff2010water,gregorio2023air,antolik2024formation}, we demonstrate that slowly evolving surface waves principally influence the splash curtain dynamics. Further work may also explore the influence of local interface shape due to waves on added-mass induced slamming forces in this axisymmetric setting, with the impactor curvature relative to the interface curvature recently shown to play a dramatic role for bluff bodies \citep{belden2024water}.  Our study complements other recent water-entry work where the interface has been disturbed from flat by other means, such as applying bulk rotation \citep{yi2021water} or via the impact of a preceding solid body \citep{rabbi2021impact}.

In conclusion, through controlled parametric experiments and scaling analysis, we have isolated the role of surface waves on cavity formation to the geometric influence of the perturbed surface at impact.  Our findings demonstrate that surface waves can alter the maximum air cavity volume by modulating the surface closure dynamics rather than the cavity expansion rate. This mechanism is further supported by the localization of the hydrodynamic perturbation due to surface waves to the near-surface region, and the large separation of timescales between the wave period and sphere entry.  
Motivated by this physical picture, this complex behavior is captured by a single non-dimensional parameter that characterizes the instantaneous wave slope at the scale of the impactor, $m\sim AR/\lambda^2$, which collapses our experimental data across a wide range of impact phases $\varphi$, frequencies $f$, and driving amplitudes $A_\textrm{d}$. Combined with the classical Weber and Bond numbers, this single parameter provides a quantitative bridge between the well-studied water-entry dynamics in idealized quiescent baths and the lesser-chartered waters of wavy interfaces.

\section*{Supplementary material.} The data and supplemental movies associated with this manuscript are openly available at \url{https://github.com/harrislab-brown/WavyWaterEntry}.

\section*{Acknowledgements.}
We gratefully acknowledge the financial support of the Office of Naval Research (ONR N00014-21-1-2816) and high-speed cameras through the Air Force Office of Scientific Research (AFOSR DURIP FA9550-24-1-0088). 
C.T.G. gratefully acknowledges support through the Hope Street Fellowship. The authors also thank Ian Gonsher for materials contributed to the custom wavemaker.

\section*{Declaration of Interests.}
The authors report no conflict of interest.

\section*{Data accessibility.}
All data reported in this study is available at \url{https://github.com/harrislab-brown/WavyWaterEntry}.

\appendix

\section{Wavemaker Description}\label{appA}

\begin{figure}
\centering
\includegraphics[width=\textwidth]{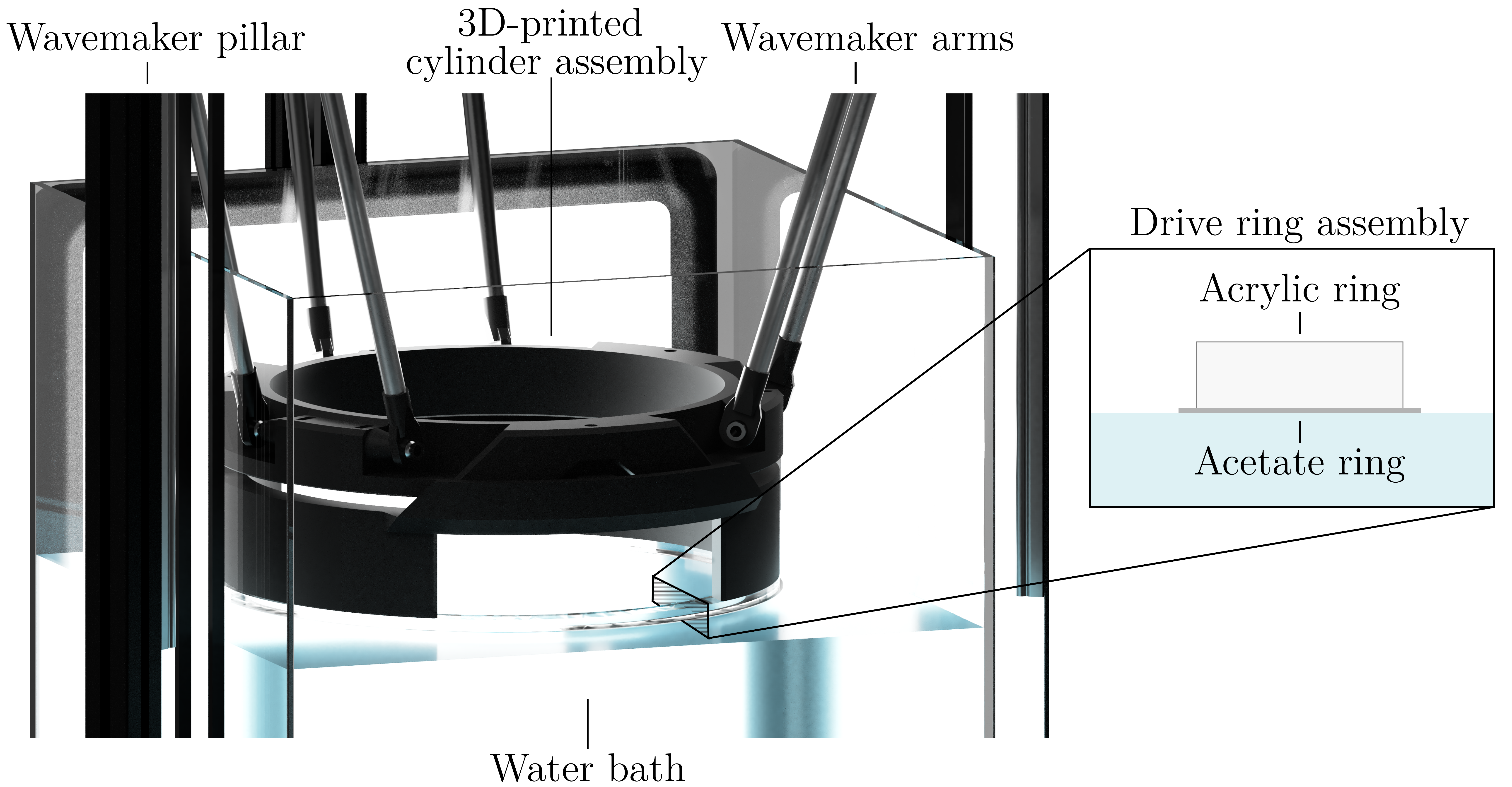} \caption{Rendering of the heaving cylindrical wavemaker.} \label{Fig6}
\end{figure}

The custom heaving axisymmetric wavemaker used in this study is based on a modified Kossel Delta 3D-printer, and is shown in detail in Figure~\ref{Fig6}.  The outer frame is constructed of aluminum extrusion consisting of three vertical pillars of height 59.2 cm, which are connected at the top and bottom to triangular bases of side length 28.5 cm. Each vertical pillar has a small carriage mounted to linear motion rail.  The carriage is connected to a belt, each of which is independently driven by a high-resolution stepper motor (17HM19-2004S, 0.9$^{\circ}$ full step size). The motors are independently driven using TMC2225 motor drivers powered by a 12 V DC power supply, and controlled by a STM32 microcontroller.  Microstepping is enabled (1/8) with each microstep corresponding to a linear carriage displacement of 6.25 $\mu$m.  Each motor is synchronously driven toward a sinusoidal target position with user-specified frequency and amplitude. The target position updates at 6.67 kHz and the resulting position error is discretized into step pulses by a phase-accumulator digital differential analyzer (DDA) algorithm running at 20 kHz. The full system is rigidly mounted to a vibration isolation table and the frame is leveled prior to experiments.

The original 3D-printer includes three identical arm pairs that extend from each carriage, and were repurposed for the setup herein.  It was found that adding a soft rubber O-ring across each set of parallel arm beams mitigated spurious high-frequency vibrations of the system.  The three arms are connected to a custom 3D-printed cylinder assembly (see Figure~\ref{Fig6}), that has three spring-loaded leveling screws for precision leveling of a lower drive ring.  A thin (0.25 mm) laser-cut acetate ring (inner diameter 13 cm, outer diameter 15 cm) is concentrically aligned and directly adhered to a laser-cut acrylic ring (thickness 3 mm). The acetate sheet extends slightly beyond the acrylic ring on both sides thereby ensuring the contact line remains strictly pinned on the sharp edges of the sheet.  This composite drive ring assembly is also concentrically aligned and adhered to the base of the 3D-printed assembly.  To ensure a consistent meniscus profile between experiments and avoid wave amplitude variability \citep{shao2021role}, the ring is positioned by lowering it in 0.1 mm increments until its bottom surface wets evenly.  This configuration and protocol ensured repeatable wave production for each set of drive parameters. 

\section{Influence of $\We$ on cavity pinchoff metrics}\label{appB}

\begin{figure}
\centering
\includegraphics[width=\textwidth]{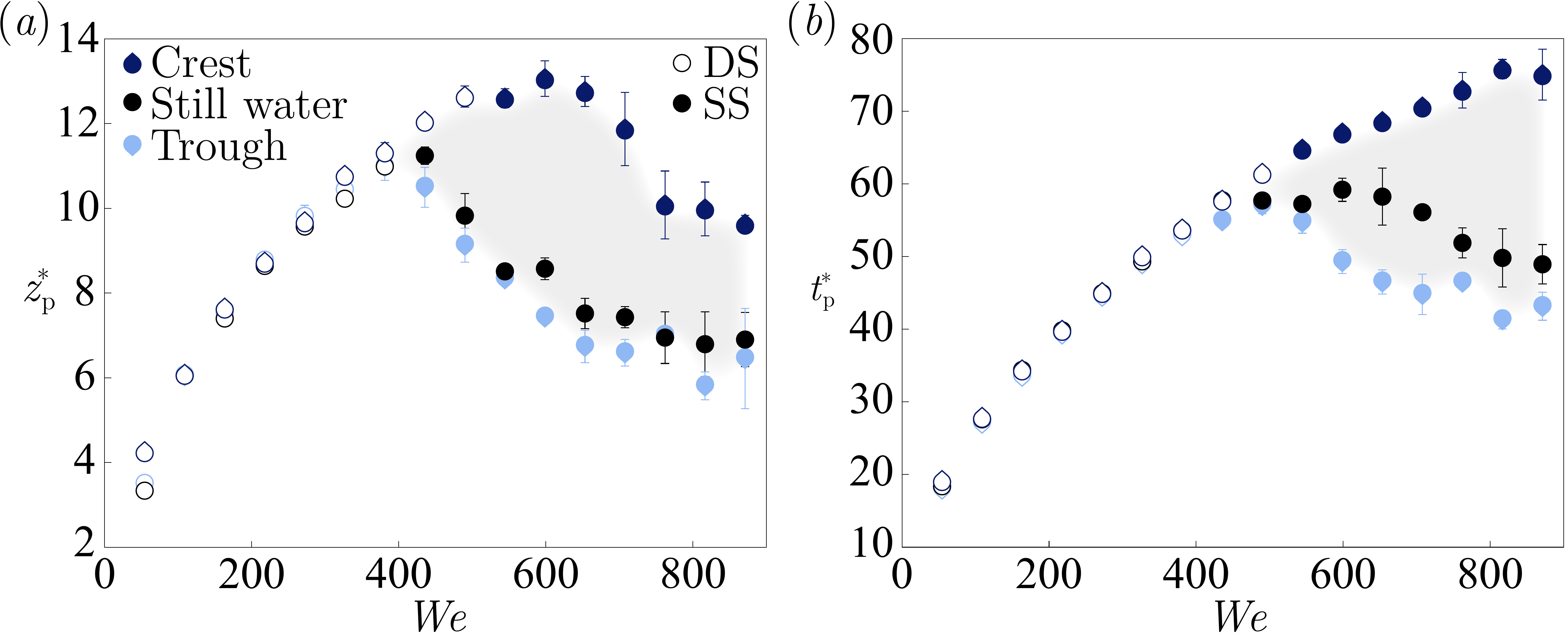} \caption{(a) Dimensionless pinchoff depth $z_\textrm{p}^{*}=z_\textrm{p}/R$ and (b) pinchoff time $t_\textrm{p}^{*}=t_\textrm{p}U/R$ against Weber number $\We$ for impacts on still water (black circles; Movie 2), a wave crest (dark blue upward cusped circles; Movie 3), and a trough (light blue downward cusped circles; Movie 4) for $A_\textrm{d}=0.1$ mm, $f=9.64$ Hz, and $\Bo=0.54$. Open and filled markers show deep seal (DS) and surface seal (SS), respectively. The shaded region highlights the Wavy Water Effect (WWE) for this set of parameters.} \label{Fig7}
\end{figure}

The influence of surface waves on maximum cavity volume first introduced in Figure~2, also influences other canonical cavity pinchoff metrics. Figure~\ref{Fig7} plots the dimensionless ($a$) pinchoff depth $z_\textrm{p}^*=z_\textrm{p}/R$ and ($b$) pinchoff time $t_\textrm{p}^*=t_\textrm{p} U/R$ against $\We$ for impacts on still water (Movie 2), wave crests $\varphi=\pi/2$ (Movie 3) and wave troughs $\varphi=3\pi/2$ (Movie 4) for $A_\textrm{d}=0.1$ mm, $f=9.64$ Hz and $\Bo=0.54$. The open and filled markers show whether the cavity closed via a deep seal or surface seal, respectively. The data follow trends similar to those discussed in the main text. Most notably, all data follow a similar increasing trend up to $\We \approx 400$, beyond which the trends depend more sensitively on the local interface geometry.

\section{Dimensionless collapse for pinchoff depth $z_\textrm{p}^*$ and time $t_\textrm{p}^*$} \label{appC}

\begin{figure}
\centering
\includegraphics[width=\textwidth]{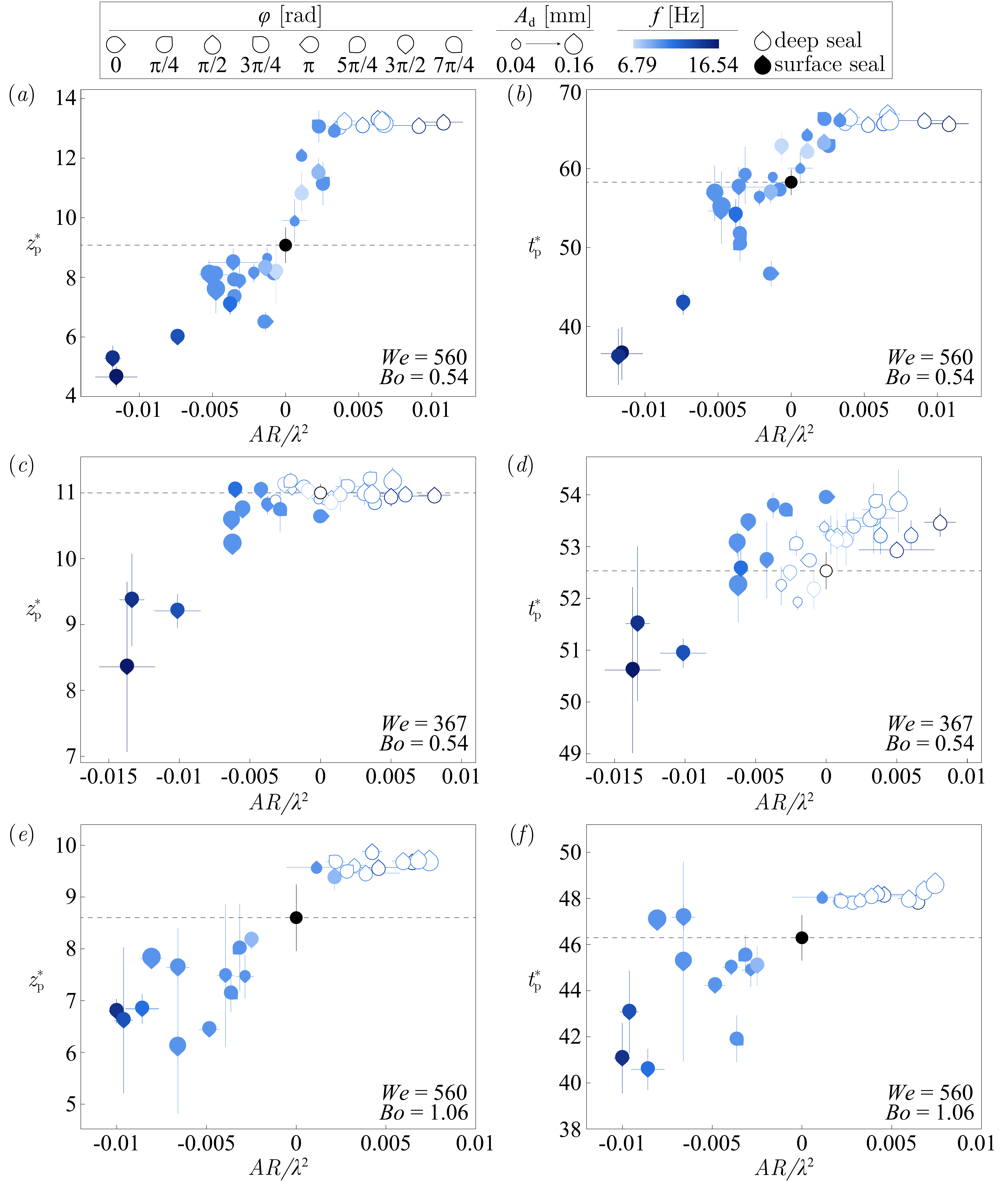} \caption{Dimensionless pinchoff depth $z_\textrm{p}^*=z_\textrm{p}/R$ and pinchoff time $t_\textrm{p}^*=t_\textrm{p}U/R$ plotted against the local wave slope $m \sim AR/\lambda^2$ for ($a,b$) $\We=560$ and $\Bo=0.54$, ($c,d$) $\We=367$ and $\Bo=0.54$, and ($e,f$) $\We=560$ and $\Bo=1.06$. The marker shape, size, and color reflect the wave phase $\varphi$, driving amplitude $A_\textrm{d}$, and driving frequency $f$, respectively, as defined in the legend. Open and filled markers denote deep seal and surface seal closure types, respectively, while the black circle represents impact on still water for identical conditions.} \label{Fig8}
\end{figure}

The dimensionless parameter $m\sim AR/\lambda^2$ characterizes the wave slope at the scale of the impacting body, defined here by the sphere radius $R$. This parameter, alongside $\We$ and $\Bo$, collapses the maximum cavity volume $V^*$ and cavity metrics at pinchoff ($z_\textrm{p}^*$, $t_\textrm{p}^*$). In Figure~\ref{Fig8}, we show pinchoff depth $z_\textrm{p}^*$ ($a,c,e$) and pinchoff time $t_\textrm{p}^*$ ($b,d,f$) for a fixed ($\We$, $\Bo$) pair plotted against $AR/\lambda^2$, where ($\We=560$, $\Bo=0.54$) for ($a,b$), ($\We=367$, $\Bo=0.54$) for ($c,d$), and ($\We=560$, $\Bo=1.06$) for ($e,f$). In all panels, the still-water reference is designated by a black circle, with its corresponding baseline metric indicated by a horizontal dashed line. For impacts resulting in a surface seal, a clear positive trend between the cavity metric and $AR/\lambda^2$ exists. This trend weakens once the impacts result in deep seals beyond a critical $AR/\lambda^2$. The strength of these trends, the critical $AR/\lambda^2$, and where the still-water reference values depend on $\We$ and $\Bo$. We also note that, as discussed above, the larger standard deviation observed in the surface seal regime reflects the disruptive influence of the downward-rushing liquid jet, which is generated during surface closure and dynamically perturbs the cavity dynamics near pinchoff.

\bibliographystyle{jfm}
\bibliography{WWE_JFM}

\end{document}